\documentclass{aadebug}
\usepackage{subfigure}
\corr{0309027}{273}

\def\reffig#1{figure~\ref{#1}}
\def\refsect#1{section~\ref{#1}}
\def\isIso(#1,#2,#3){$\langle${\tt x}=#1,{\tt y}=#2,{\tt z}=#3$\rangle$}

\begin{document}

\runningheads{Brock Pytlik et al.}{Automated~Fault~Localization
Using Potential Invariants}

\title{Automated Fault Localization Using Potential
Invariants\footnotemark}

\author{Brock~Pytlik,
Manos~Renieris\footnotemark, 
Shriram~Krishnamurthi and Steven~P.~Reiss%
}

\def\address#1#2{\parbox{32pc}{\raggedright\small\textit{#2}}}

\address{1}{%
Brown University,
Computer Science Department,
Providence, RI 02912,
USA%
}

\extra{1}{This work is partially supported by the National
  Science Foundation.}
\extra{2}{Contact email: \texttt{er@cs.brown.edu}}

\pdfinfo{
/Title (Automated Fault Localization Using Potential Invariants)
/Author (Brock Pytlik, Manos Renieris, Shriram Krishnamurthi and Steven P. Reiss)
}

\begin{abstract}
  We present a general method for fault localization based on
  abstracting over program traces, and a tool that implements the
  method using Ernst's notion of potential invariants.
  Our experiments so far have been unsatisfactory, suggesting
  that further research is needed before invariants can be used to
  locate faults.
\end{abstract}


\section{Introduction\label{sec:intro}}

Suppose a programmer receives a report of a bug in a thoroughly tested
program.  The report takes the form of an input on which the
programmer expects the program to work correctly.  This paper
describes a generic method, and a corresponding tool, to help
programmers in this situation.  The method requires the following
three pieces of information:
\begin{itemize}
\item
a source program, to instrument;
\item
an input that exposes a bug (the ``bad input''); and,
\item
a set of inputs on which the program executes successfully (``good
inputs'').
\end{itemize} 
The first two are natural requirements for debugging.  The assumption
that the program has been  thoroughly tested guarantees the existence
of good inputs.

Given this information, the method
\begin{itemize} 
\item
instruments the program to garner execution traces;
\item
produces a \emph{spectrum}, i.e., a concise description of the
program's behavior, of each input's execution;
\item
combines the spectra of the successful runs to generate a single,
summary \emph{model\/} of their behavior; and,
\item
contrasts this model with the spectrum from the failing run and
reports the difference.
\end{itemize}
If the spectra of the runs and the model of the good runs are accurate
enough, the difference will reflect the bug.  This paper presents
preliminary experiments that evaluate this method for a particular
category of spectra, namely potential invariants.

\section{Debugging in Principle: An Example\label{sec:invariants}}

\newcommand{\lte}[2]{\texttt{#1}$<$\texttt{#2}}

\begin{figure}[t]
  \subfigure[Sample Program\label{fig:isiso}]{%
    \begin{minipage}{0.5\textwidth}
      \texttt{%
        bool isIsosceles (x, y, z) \{ \\
        \hspace*{2pc}      if (x == y) return 1; \\
        \hspace*{2pc}      else if (y == z) return 1;\\
        \hspace*{2pc}     else return 0;              
        \} } \end{minipage}
  }%
  \subfigure[Live Invariants for \texttt{isIsosceles}\label{fig:geninv}]{
    \begin{minipage}{0.5\textwidth}
    \begin{tabular}{cc}
      after \isIso(1,2,3): & 
      \lte{x}{y}, \lte{x}{z}, \lte{y}{z}
      \\
      after \isIso(2,5,5): & 
      \lte{x}{y}, \lte{x}{z}
      \\
      after \isIso(2,2,3): &
      \lte{x}{z}
    \end{tabular}
   \smallskip\smallskip
    \end{minipage}
  }
  \hrule
\caption{Sample program and live invariant sets.}
\end{figure}

Potential invariants, developed by Ernst et al.~\cite{DaikonTSE},
relate variables at static program locations.  A ``live'' invariant is
one that has not yet been falsified by a run.  To tractably identify
potential invariants, a tool must limit them to a fixed set of
schemata.

Consider the function \verb|isIsosceles| in \reffig{fig:isiso}, and
suppose we have only one invariant schema: \lte{$a$}{$b$} where $a$ and $b$
are metavariables.  Instantiating this schema at the entry of
\verb|isIsosceles| yields six concrete potential invariants:
\begin{center}
\lte{x}{y} \quad
\lte{x}{z} \quad
\lte{y}{x} \quad
\lte{y}{z} \quad
\lte{z}{x} \quad
\lte{z}{y}
\end{center}
Each line in the table in
\reffig{fig:geninv} shows the set of invariants that have not been
falsified by all the runs preceding and including that line.
The last line therefore shows the potential invariants that survived
all three executions of the function.  

Observe that these three inputs do not expose the error in the
function, which is that it fails to compare \verb|x| with \verb|z|.
Now suppose we are given an input that does expose the error, such as
\isIso(2,3,2).
For the bad input, the set of live potential invariants after only this input is
\begin{center}
\lte{x}{y} \quad
\lte{z}{y}
\end{center}

We can use the potential invariants for any input (starting with all
six potential invariants) as its spectrum of the run with that input.
We must now contrast the spectrum of the bad run with a model for the
successful runs.  The set of invariants left at the end of the three
good runs, i.e. the intersection of their spectra, constitutes a
plausible such model.  Computing the set difference between the
model of the good runs and the spectrum of the bad run produces
\begin{center}
\lte{x}{z}
\end{center}
This invariant clearly demonstrates that the program relies on the
condition \lte{x}{z} being true, which the fault-inducing input violates.

\section{Debugging in Practice\label{sec:impl}}

We have developed a program, Carrot, that implements the debugging
technique presented in \refsect{sec:invariants}.  The potential
invariants, in the style of Daikon~\cite{DaikonTSE}, are drawn from
the following relational schemata, instantiated on function entries
and exits:
\begin{itemize}
\item 
The \emph{equality\/} invariant checks whether two variables are
always equal.
\item
The \emph{sum\/} invariant checks whether two variables always sum to
the same value.
\item
The \emph{less than\/} invariant checks whether a variable is always
less than another variable.
\item
The \emph{constant equality\/} invariant checks whether a variable is
always equal to a constant.
\end{itemize}
In addition, Carrot generates \emph{value sets}, which record the set
of all values bound to a variable.  In contrast to potential
invariants, which are falsified by runs, value sets are initially
empty and acquire values over the course of execution.

This simple form of value sets, however, can lead to needless
inaccuracy in estimating program behavior.  Suppose a function $f$ has
two formal arguments, $x$ and $y$.  If $f$ is called twice, once on
$\langle1,3\rangle$ and the second time on $\langle2,4\rangle$, the value set for $x$
contains $1$ and $2$, and the value set for $y$ contains $3$ and $4$.
The cross-product of these sets contains the pair $\langle1,4\rangle$,
which incorrectly suggests that the call $f(1,4)$ occurred in the
program's execution.  To diminish this form of inaccuracy, we also
maintain sets of \emph{pairs of values} that occurred during
execution.

Carrot uses Daikon's instrumenter to annotate programs, then executes
them to generate traces.  It analyzes these traces to compute spectra.
Carrot then generates the model from these spectra, and then computes
the difference between the model and spectrum for the bad run.
Specifically, Carrot incorporates the bad run's spectrum into the
model and observes which components of the model change.

It appears to be possible to employ Daikon to compute these spectra
and models through a judicious use of command-line flags (to, for
instance, disable Daikon's use of confidence levels in reporting
potential invariants).  Future experiments should therefore use Daikon
for this purpose, so they may exploit its wider range of invariant
schemata.\footnote{Note, however, that Daikon currently does not
support pairs of value sets.}

\section{Experiments\label{sec:experiments}}

To evaluate Carrot's effectiveness, we must determine whether
\begin{enumerate}

\item
the model eventually converges to a ``steady state'', i.e., additional
good runs do not significantly alter it; 

\item
contrasting the model of good runs with the spectrum for a bad run
produces anything at all; and,

\item
the difference holds a clue to the actual error.

\end{enumerate}
Comparing a premature model of good runs against a bad run can result
in the tool reporting many more potential invariants than comparing with
a steady state model.  The first item
is therefore important for minimizing the number of invariants the
programmer would need to examine.

We tested Carrot on two programs in the Siemens
suite~\cite{hutch,Harrold:spectra}:
\begin{itemize}
\item
\emph{tcas} (140 lines of code) implements an airplane collision
avoidance decision
process.  It is essentially a large predicate on 13-tuples of
integers. 
\item
\emph{print\_tokens} (615 lines of code) tokenizes its input file and outputs the set of
tokens.
\end{itemize}
For each program there are a correct version, a number of versions with
a single inserted fault (41 for \emph{tcas}, 7 for
\emph{print\_tokens}), and a large set of inputs (1592 inputs for
\emph{tcas}, 4072 inputs for \emph{print\_tokens}).  Each faulty
version manifests its error on at least one input.

To check the eventual stability of the model, we experimented with the
correct versions of \emph{tcas} and 
\emph{print\_tokens}\footnote{%
  Further experiments with a third program from the Siemens suite,
  \emph{replace}, which performs regular expression substitution in
  strings, produced similar results.}.  The results are different for
value sets and relational invariants.  We find that the size of the
set of relational invariants decreases rapidly, and eventually reaches
a steady state.  In contrast, even the last five runs of the programs
cause hundreds of value set extensions.

To check the second and third requirements, we examine all faulty
versions available for \emph{tcas} and \emph{print\_tokens}.  We found that contrasting the
faulty run to the steady state model does not always result in
invalidations.  In particular, there is no invalidation for any of the
568 faulty runs of versions of \emph{tcas}.  For \emph{print\_tokens},
out of 484 faulty runs, only one (of version~5) invalidates relational
invariants. The two invalidated invariants are not related to the bug.

We then created a new faulty version of \emph{print\_tokens}, with the
explicit purpose of uncovering the bug in it.  \emph{print\_tokens}
contains a partial identity function, which returns its argument if
the argument is in a specific set of values, and terminates the
program otherwise. In our version, the function returns its argument
if the argument is in the set of values, otherwise (to inject a fault)
it returns the largest value of the set.  The result is that the
function is not an identity function anymore. This bug should be identified
by the invariant schemata we implemented.
  
Our manufactured version of \emph{print\_tokens} exposes its bug on 48
inputs.  Each faulty run invalidates the same two invariants, which
point directly to the bug, except one run which invalidates two more
invariants, which are unrelated to the bug.

\section{Conclusions\label{sec:future}}

Our very preliminary experience based on this work is naturally
negative.  We were unable to realistically locate any bugs in the many
variants of the Siemens suite.  This failure could be caused by any
number of shortcomings in our approach: Carrot doesn't implement a
sufficiently rich set of invariants; the style of invariants used by
Carrot is mismatched with the programs we're analyzing; or potential
invariants are not suitable for debugging.  Our experiments are too
premature to conclude the last point.  On the one hand, the potential
for success clearly exists, as the manufactured version of
\emph{print\_tokens} suggests, and as  tools more closely
hewn to a particular domain, such as Diduce~\cite{diduce},
demonstrate.  On the other hand, a similar technique used by Groce and
Visser~\cite{GroceSPIN03} has not been shown to bear
fruit.\footnote{Visser, personal communication.}

\section*{Acknowledgment}
Michael Ernst carefully read a draft of this paper and provided many
helpful suggestions.

\bibliography{local}

\begin{thebibliography}{HRWY98}

\bibitem[ECGN01]{DaikonTSE}
Michael~D. Ernst, Jake Cockrell, William~G. Griswold, and David Notkin.
\newblock Dynamically discovering likely program invariants to support program
  evolution.
\newblock {\em IEEE Transactions on Software Engineering}, 27(2):99--123,
  February 2001.

\bibitem[GV03]{GroceSPIN03}
Alex Groce and Willem Visser.
\newblock What went wrong: Explaining counterexamples.
\newblock In {\em Proceedings of the 10th International SPIN Workshop on Model
  Checking of Software}, volume 2648 of {\em Lecture Notes in Computer
  Science}, 2003.

\bibitem[HFGO94]{hutch}
Monica Hutchins, Herb Foster, Tarak Goradia, and Thomas Ostrand.
\newblock Experiments on the effectiveness of dataflow- and controlflow-based
  test adequacy criteria.
\newblock In {\em Proceedings of the 16th International Conference on Software
  Engineering}, pages 191--200, 1994.

\bibitem[HL02]{diduce}
Sudheendra Hangal and Monica~S. Lam.
\newblock Tracking down software bugs using automatic anomaly detection.
\newblock In {\em Proceedings of the 24th International Conference on Software
  Engineering}, pages 291--301, 2002.

\bibitem[HRWY98]{Harrold:spectra}
Mary~Jean Harrold, Gregg Rothermel, Rui Wu, and Liu Yi.
\newblock An empirical investigation of program spectra.
\newblock In {\em Proceedings of the ACM SIGPLAN-SIGSOFT Workshop on Program
  Analysis For Software Tools and Engineering}, pages 83--90, 1998.

\end{thebibliography}
\end{document}